# NUMERICAL PREDICTION AND POST-TEST NUMERICAL ANALYSIS OF THE ASDMAD WIND TUNNEL TESTS IN ETW

Manuel Brüderlin[1], Norbert Hosters[1], Bae-Hong Chen[1], Alexander Boucke[2], Josef Ballmann[2], and Marek Behr[1]

[1]CATS, RWTH Aachen University
Schinkelstr. 2, 52056 Aachen, Germany
{bruederlin,hosters,chen,behr}@cats.rwth-aachen.de

[2]LFM, RWTH Aachen University and ITAM Consulting
Schinkelstr.2, 52056 Aachen, Germany
{ballmann,boucke}@lufmech.rwth-aachen.de

**Keywords:** Aero-Structural Dynamics, Steady/Unsteady Computational Aeroelasticity,

**Abstract:** This paper presents numerical results in comparison with experimental data of nominally static aeroelastic polars of the second ASDMAD (Aero-Structural Dynamics Methods for Airplane Design) campaign conducted in the European Transonic Windtunnel. For the simulation the modular solver package SOFIA combined with the DLR solver TAU on the fluid side and the in-house solver FEAFA on the structural side is applied. First the experimental setup and the aeroelastic solver are introduced, followed by a detailed analysis of the results. Comparisons of pressure distributions and the global lift coefficients are presented for variations of angle of attack, Mach number and aerodynamic loading factor. Additionally, the paper contains preliminary investigations of dynamic results for excited vibration applying periodical inner force couples in the wing root area.

## 1 INTRODUCTION

The provision of data from static and dynamic aeroelastic experiments in transonic flow regimes at high Reynolds number was the main purpose of the HIRENASD (High Reynolds Number Aero-Structural Dynamics) wind tunnel experiments. The project was part of the German Research Foundation (DFG) funded collaborative research center "Flow Modulation and Fluid-Structure Interaction at Airplane Wings" (SFB 401) and the experiments were conducted in the European Transonic Windtunnel (ETW) in November 2006 [1,2]. The measured data of this project was used for the validation of several computational aero-structural dynamics (CASD) solvers. Parts of the data have been made available for the Aeroelastic Prediction Workshop [3]. In the follow-up project ASDMAD the wing was modified with two different winglet geometries for two measuring campaigns by keeping the original projected semispan [4]. In the related wind tunnel campaigns the modified wing models were studied for varying Mach numbers and load factors q/E (q dynamic pressure, E Young's modulus of the wing model). Similar to the HIRENASD project during the quasi-static polars the aeroelastic behavior of the wing was studied at continuously but very slowly varying angles of attack. During the dynamic experiments, in both ASDMAD campaigns the piezoelectric mechanism of the HIRENASD experiments [1] was used to excite the wing for vibration at a constant angle of attack. Additionally, during the second ASDMAD campaign the winglet control surface was deflected with varying constant angles during the static polars and with a piezoelectric dynamic excitation during the dynamic polars.

In the present paper a selection of gathered data from the ASDMAD wind tunnel campaigns is compared with numerical results of aeroelastic and purely aerodynamic computations ignoring deformation. For the simulations the in-house software package Solid Fluid Interaction (SOFIA) is used. The software package combined with the DLR flow solver FLOWer predicted the equilibrium states of the HIRENASD project very well [5]. But for the present paper the DLR flow solver TAU is used instead.

## 2 THE ASDMAD 2 WIND TUNNEL MODEL

The ASDMAD 2 wind tunnel model is based on the elastic semi-span wing model from the HIRENASD Project, which is provided by the DFG funded collaborative research center "Flow Modulation and Fluid-Structure Interaction of Airplane Wings" (SFB 401). Within the HIRENASD project the wing had been designed, constructed, equipped with measurement techniques, qualified and tested in ETW [1,2]. The wing planform is typical for large transonic passenger aircrafts. The basic design, the wind tunnel configuration and the integrated excitation mechanism, which allows the application of an oscillating internal bending moment is described in detail in [1]. Within the ASDMAD project the HIRENASD wing was modified. Whilst keeping the span, the wing tip was removed and replaced with two different winglet configurations, one larger one-part winglet for the ASDMAD-1-campaign and one two-part winglet with an aerodynamic control surface (ACS) for the ASDMAD-2 campaign, which is considered in this presentation. The ASDMAD-2 winglet has a dihedral of 40° and 10° additional sweep relative to the main wing leading edge sweep. Two piezo actuators arranged in an X-Frame allow a dynamic excitation of the control surface up to 30Hz [6,7].

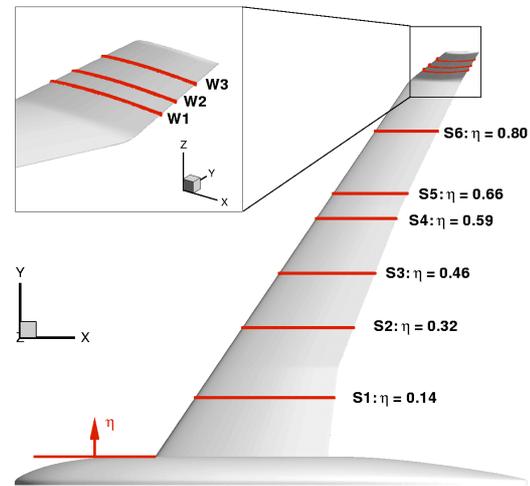

Figure 1: Pressure sections at wing and winglet

The wing is equipped with six different measurement systems, which are described in detail in [1,4]. The experimental data used for comparison in the presented paper is measured with over 200 in-situ pressure transducers (Kulites). Most of them are arranged in six pressure sections on the wing (see Figure 1). Nine pressure transducers are located in the winglet sections marked as W1, W2 and W3 in Fig. 1. In addition the balance data is used.

## 3 THE APPLIED AERO-STRUCTURAL DYNAMICS SOLVER

The applied aeroelastic solution scheme is based on a partitioned approach and is built-up as a modular software system. The Aeroelastic Coupling Module (ACM) is the key element [8,9] and its comprehensive range of interfaces provides a completely single-field independent solver. Several spatial as well as temporal coupling methods, which are needed for the data transfer between non-matching meshes and for the synchronization of the involved single-field solvers, are implemented. The application of solver-specific meshes within an aeroelastic solver requires projection methods between the different surface discretisations. The spatial transfer methods for the aerodynamic loads and structural deformations at the fluid-structure interface are required to be valid from the physical point of view. To fulfill the condition of conservation the following two criteria have to be satisfied by any projection scheme. Each projection scheme has to preserve the total force and moment vectors. During steady simulations, the work performed by the aerodynamic loads on the wetted surface must be equal to the elastic strain energy of the structure. In addition, the instantaneous power exchange over the interface must be equal on both sides during unsteady simulations. Furthermore, the volume mesh can also affect the numerical accuracy of the flow solver and the coupling strategy itself. A smooth transfer of the computed deformation to the surface as well as the volume mesh is therefore essential. The ACM provides several projection methods. In the present work the Finite Interpolation Method [10] is applied, which directly uses the shape functions of the structural model to distribute the aerodynamic loads among the nodes of the closest structural element. Further implemented methods are the Global Spline Based method [11] and the Moving Least Squares method [12].

For the temporal synchronization the ACM provides several weak and strong coupling schemes [8,9]. In the present work a strong coupling based on a Gauss-Seidel approach with a fixed relaxation is applied for the steady simulations.

To solve the structural problem, the in-house code Finite-Element Analysis for Aeroelasticity (FEAFA) is used. Besides several shell and volume elements it provides a multi-axial Timoshenko beam element [13], which allows for accurate computations of the structural displacement of slender structures at low computational costs. For unsteady computations a Bossak-Newmark scheme with subcycling is implemented. For further reductions of the computational costs during the time-integration the structural equations can be solved in modal coordinates. The 3D time-dependent Reynolds-averaged Navier-Stokes (RANS) equations for perfect gases on deformable grids are solved with the CFD solver TAU [14], which is developed under the leadership of the German Aerospace Centre (DLR). For the closure of the RANS equations this unstructured Finite-Volume solver provides several algebraic, eddy-viscosity-based and Reynolds stress turbulence models.

For the technical coupling of the whole software package the FlowSimulator software [15] is used. It was developed by Airbus, EADS Military Air Systems and several European research institutes to offer a common platform for efficient multi-disciplinary solvers on high-performance computers. The FSDataManager (FSDM) as the central module provides a massively parallel in-memory data storage, which allows for a highly efficient data exchange between connected modules, especially single-field solvers. Besides the FSTau module, which encapsulates the TAU solver, a mesh deformation tool based on radial basis functions is offered [16]. This tool is used for the deformation of the volume meshes in the present work.

# 4 RESULTS

## 4.1 Computational Setup

The applied beam model (see Figure 2) consists of about 600 Timoshenko beam elements. Besides the wing it also includes clamping, excitation mechanism, balance and the ETW adapter, since these parts severely influence the modeshapes and eigenfrequencies of the beam model [5]. For the solution of the structural dynamics system of equations a modal approach with 24 modes is used. Due to the different temperature levels in the ETW for different flow conditions the Young's-modulus and shear-modulus of the beam model are adjusted accordingly. The control surface deflection is idealized as smooth additional prescribed surface displacement around its rotation axis, as shown in Figure 2.

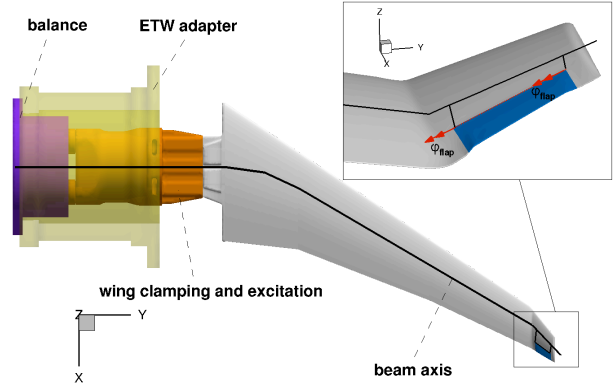

Figure 2: View of the test assembly including the representative elastic beam and ACS deflection

The unstructured CFD Mesh has about 7.26 million grid points and was generated with ANSYS ICEM CFD. In the vicinity of wing and fuselage where the no-slip boundary condition is applied, prism layers are used to accurately resolve the boundary layer. The initial height of the prism layer was chosen so that y+ is about 1. The wall on which the wing is mounted is represented by a symmetry boundary condition neglecting viscosity. For the remaining surfaces a farfield boundary condition is applied. Owing to the high Reynolds number of 23.5 million in the experiment, the no-slip walls are considered fully turbulent. For turbulence modeling the one equation Spalart-Allmaras modell with Edwards modification (SAE) [17] is applied.

## 4.2 Results

### 4.2.1 Aeroelastic Behaviour with Respect to the Angle of Attack

All shown CFD-CSM computations are in very good agreement with the experimental data, especially on the whole pressure side and behind the shock on the suction side. In front of the shock an almost constant offset is visible. With increasing angle of attack the shock on the suction side of pressure section 6 moves towards the leading edge and increases in strength. At $\alpha=4.5°$ the shock induces a massive flow separation as indicated by the wide value distribution of measured cp data. Even if the RANS-equations cannot capture the value distribution of cp they still fit quite accurately in the maximum probability dots of the experimental data.

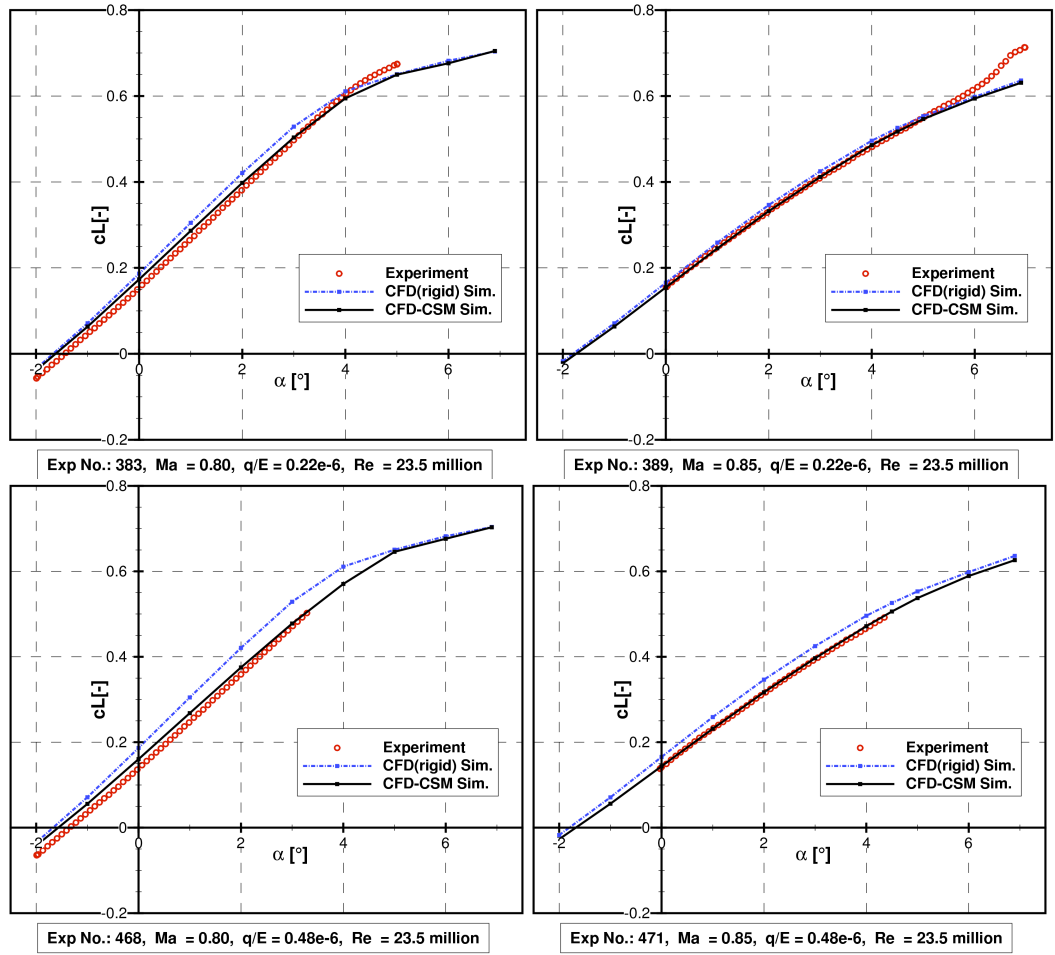

Figure 3 shows a comparison of the global lift coefficient over the angle of attack between CFD-CSM simulation results and experimental data. Additionally, computational results for a rigid assembly are depicted to evaluate the benefit of the CFD-CSM coupling procedure. The ACS is locked at -5° (upwards/towards the fuselage) for all steady conditions compared in this paper. The left column shows Ma=0.8 and the right Ma=0.85 while the loading factor q/E increases from top (q/E=0.22·10$^{-6}$) to bottom (q/E=0.48·10$^{-6}$).

Due to the kinematic coupling of flap bending angle and angle of attack (AoA) for backwards swept elastic wings, an increasing bending in spanwise direction contributes, besides torsion, to the aerodynamic twist with decreasing the relative angle of attack and, therefore, to a lower outboard loading in comparison to rigid configurations. Together with the torsional deformation, this effect causes mainly the slope difference between the flexible and rigid computations while the flow is attached. With increasing flow detachment at the higher angles of attack, the difference between the rigid and flexible computations decreases again. Due to the stronger flow separation the linear region of cL($\alpha$)-is much smaller at Ma=0.85.

All CFD-CSM computations show a much better agreement with the experimental data than the rigid wing CFD computations. At Ma=0.8, there is still a slight slope difference between CFD-CSM and the experimental data whereas at Ma=0.85 the agreement is very good in the range of measured data except for $\alpha > 6°$ at q/E=0.22·10$^{-6}$.

Figure 4 depicts the changes in the pressure distribution in section 6 due to an increasing angle of attack at Ma=0.85, q/E=0.48·10$^{-6}$. Besides the comparison between experimental data and CFD-CSM results, rigid wing CFD results are also shown. In addition to the maximum probability of recorded experimental cp data the value distribution at each pressure sensor is shown. For the sake of clarity the suction side is shown with a negative x/c.

All shown CFD-CSM computations are in very good agreement with the experimental data, especially on the whole pressure side and behind the shock on the suction side. In front of the shock an almost constant offset is visible. With increasing angle of attack the shock on the suction side of pressure section 6 moves towards the leading edge and increases in strength. At α=4.5° the shock induces a massive flow separation as indicated by the wide value distribution of measured cp data. Even if the RANS-equations cannot capture the value distribution of cp they still fit quite accurately in the maximum probability dots of the experimental data.

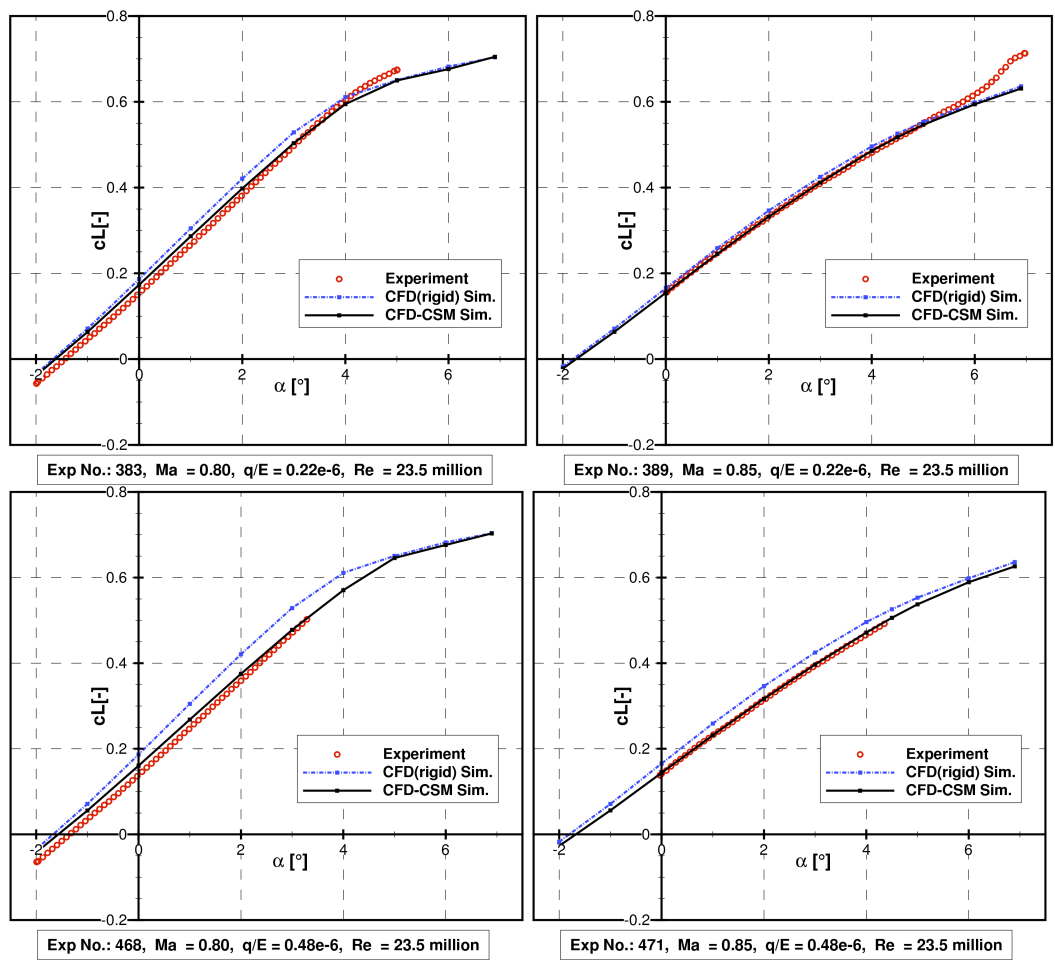

**Figure 3: Global lift coefficient over angle of attack for different Mach numbers (left: Ma=0.8, right: Ma=0.85) and load factors q/E=0.22·10$^{-6}$ (top) and 0.48·10$^{-6}$ (bottom)**

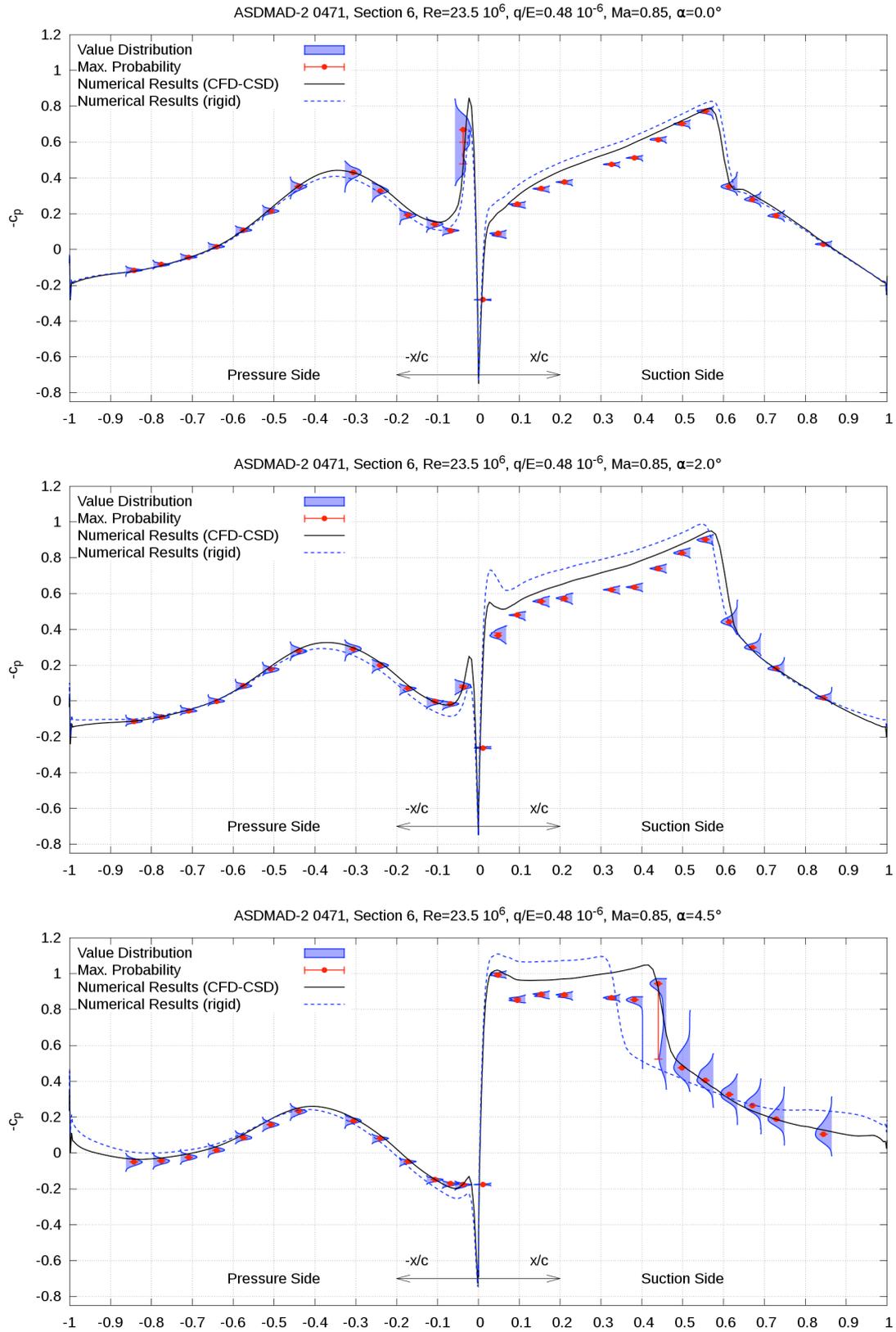

**Figure 4: Pressure distribution in section 6 for different angles of attack $\alpha=0°$ (top), $2°$ (middle) and $4.5°$ (bottom), Ma=0.85, q/E=0.48·$10^{-6}$ (Exp. ASDMAD-2 471)**

### 4.2.2 Mach number effect

With increasing Mach number the shock strength increases as well as the shock induced flow separation. This leads to a decrease in cL, as depicted in Figure 5 for various angles of attack at q/E=0.34·10$^{-6}$. Due to the lower lift the differences between elastic wing and the rigid wing computation decrease. The agreement between computational and experimental data is very good for Ma > 0.83. At Ma=0.8 the CFD-CSM results are slightly lower for α < 3° and slightly higher for α > 3°.

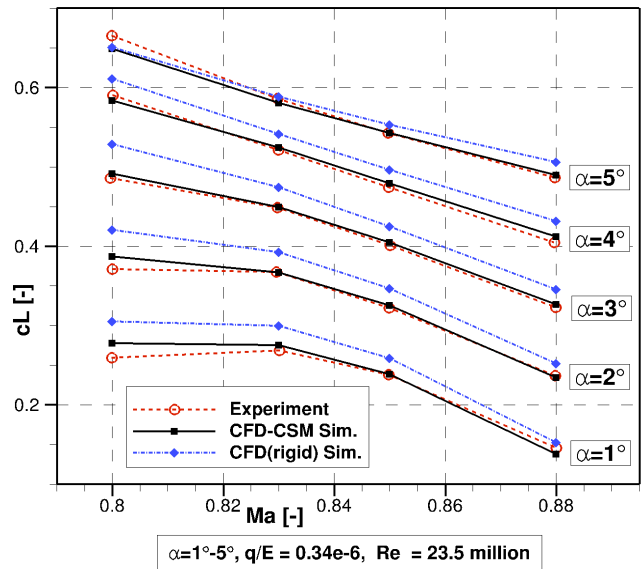

Figure 5: Global lift coefficient over Mach number for different angle of attacks, α=1°-5°, Ma=0.80-0.88, q/E=0.34·10-6

Figure 6 depicts a Mach number variation of the pressure distribution in pressure section 6 at α=2° and q/E=0.34·10$^{-6}$. Despite the higher computed pressure level in front of the shock on the suction side the agreement with the experimental data is very good. The lift reduction with increasing Mach number due to the increasing shock strength is clearly visible. At Ma = 0.88 the value distribution of the experimental data indicates a strong shock induced flow separation. In this case the computational result does fit only purely in the dots of the most probable values of the experimental data.

### 4.2.3 Aeroelastic effects on spanwise loading

The lower outboard loading due to the aerodynamic twist of swept elastic wings in comparison to rigid configurations is clearly visible in Figure 7 and Figure , which present the experimental and numerical pressure distributions for two wing sections and one winglet section for M=0.85, q/E=0.48·10$^{-6}$ and α=2°. While the pressure distribution of the most inboard section is almost equal for the rigid and flexible computation, a clear difference can be observed for the fourth, the sixth and the winglet section. Here the pressure level is lower for the flexible configuration and the shock position on the suction side is located towards the trailing edge. Especially the pressure level of the numerical results of the flexible computation is in a good agreement with the experimental data for the suction as well as the pressure side. For the fourth section a shock-induced flow separation can be expected on the suction side, which is confirmed by the width of the experimental data distribution. Although the most probable pressure is in good agreement with the numerical results, this leads to the assumption of unsteady effects, which cannot be captured by the steady RANS equations. For the outboard sections the unsteady effects disappear, which has been also observed for the HIRENASD wing under comparable flow conditions. In general, a clear improvement of the flexible computations in comparison to the rigid ones is obvious.

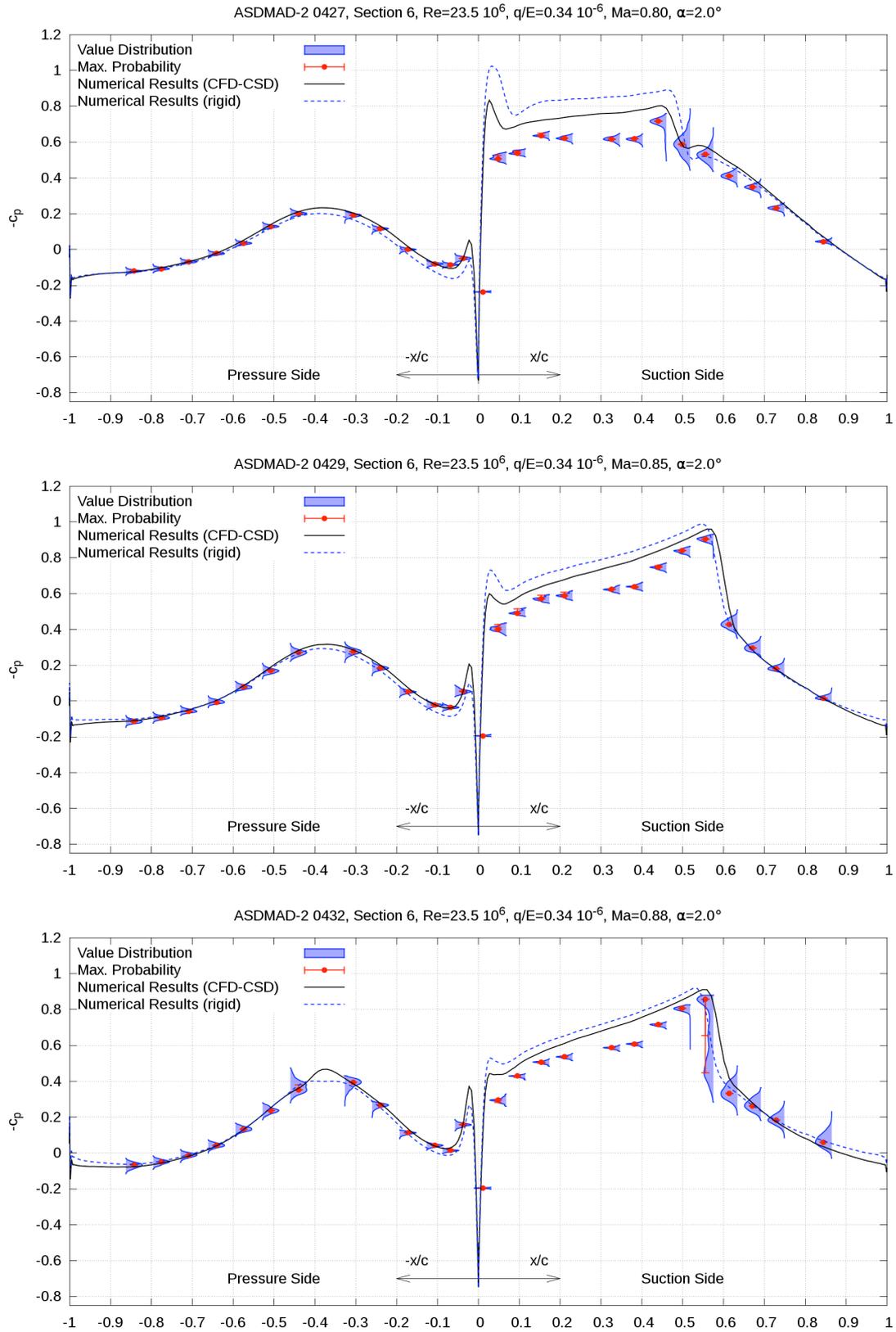

**Figure 6: Changes of pressure distribution with increasing Mach number in section 6 for $\alpha$=0°, q/E=0.34·$10^{-6}$, Ma=0.8 (top), 0.85 (middle) and 0.88 (bottom)**

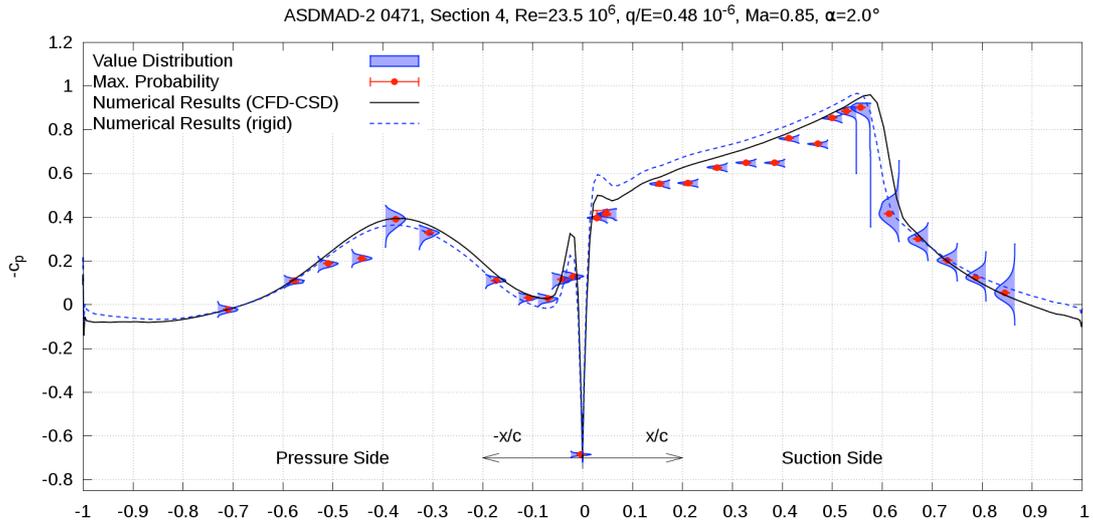

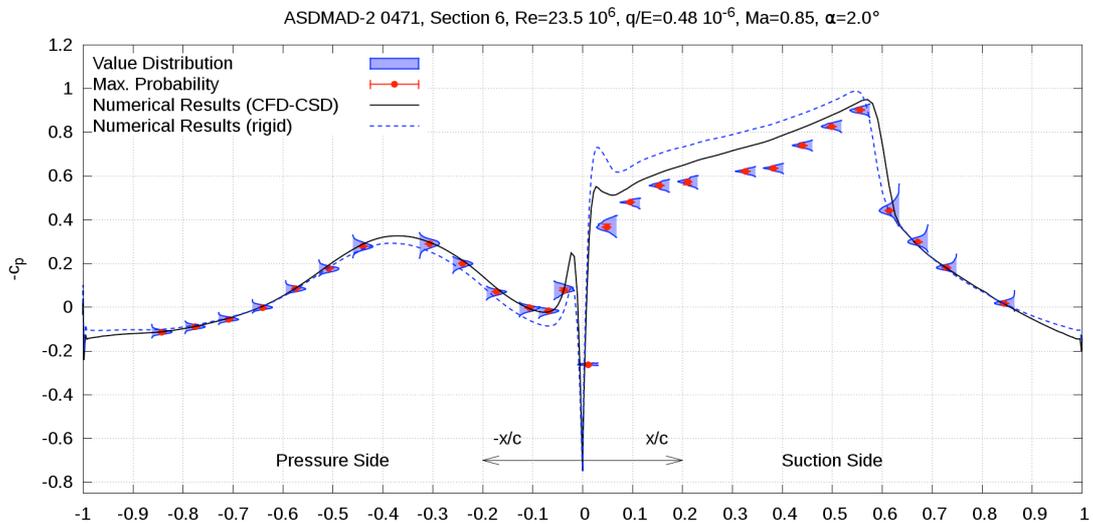

**Figure 7:** Aeroelastic effects on spanwise loading in sections 1, 4 and 6 for $\alpha=2°$, $q/E=0.48 \cdot 10^{-6}$, Ma=0.85

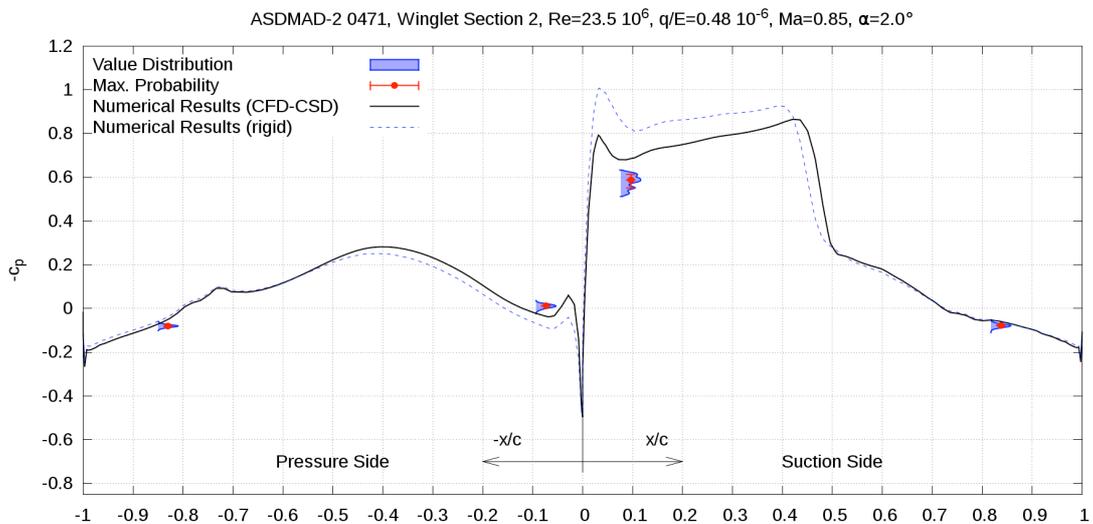

**Figure 7:** Aeroelastic effects on spanwise loading in winglet section 2, for $\alpha=2°$, $q/E=0.48 \cdot 10^{-6}$, Ma=0.85

### 4.2.4 Influence of the Increasing Load Factor

While the dynamic pressure does not affect the aerodynamic behavior of rigid configurations, flexible configurations are significantly influenced. The increasing loading factor q/E causes higher deformation and thus a drop of the outboard loading of the backwards swept wing. This is confirmed by the pressure distributions for the most outboard section presented for M=0.85, α=4.5° and two loading factors q/E (0.22·10$^{-6}$, 0.48·10$^{-6}$) in Figure 9. On the one hand, the local lift level is decreasing with increasing loading factor, while the pressure distribution of the rigid computations remains constant. On the other hand, the shock position is moving to the trailing edge. As observed in the previous sections, the increased width of the experimental data distribution behind the shock points to a strong flow separation. But at least, the numerical results of the flexible computations match the most probable experimental pressure quite well. In addition to these local effects a large impact to the global coefficients is evident. This is confirmed by the results depicted in Figure 8, where the cL over the q/E is presented for M=0.85 and different angles of attack. Combined with the increasing loading caused by an increasing angle of attack, the gradient of the cL(α)-curve decreases. This effect was also observed in All shown CFD-CSM computations are in very good agreement with the experimental data, especially on the whole pressure side and behind the shock on the suction side. In front of the shock an almost constant offset is visible. With increasing angle of attack the shock on the suction side of pressure section 6 moves towards the leading edge and increases in strength. At α=4.5° the shock induces a massive flow separation as indicated by the wide value distribution of measured cp data. Even if the RANS-equations cannot capture the value distribution of cp they still fit quite accurately in the maximum probability dots of the experimental data.

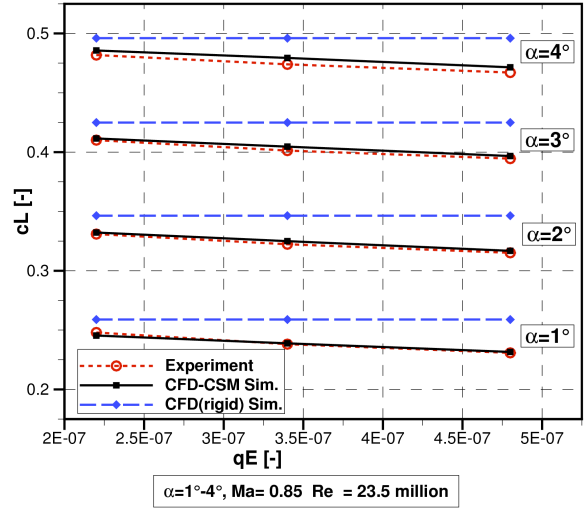

**Figure 8: Global lift coefficient over the load factor q/E for different angle of attacks, α=1°-4°, Ma=0.85, q/E=0.22·10$^{-6}$- 0.48·10$^{-6}$**

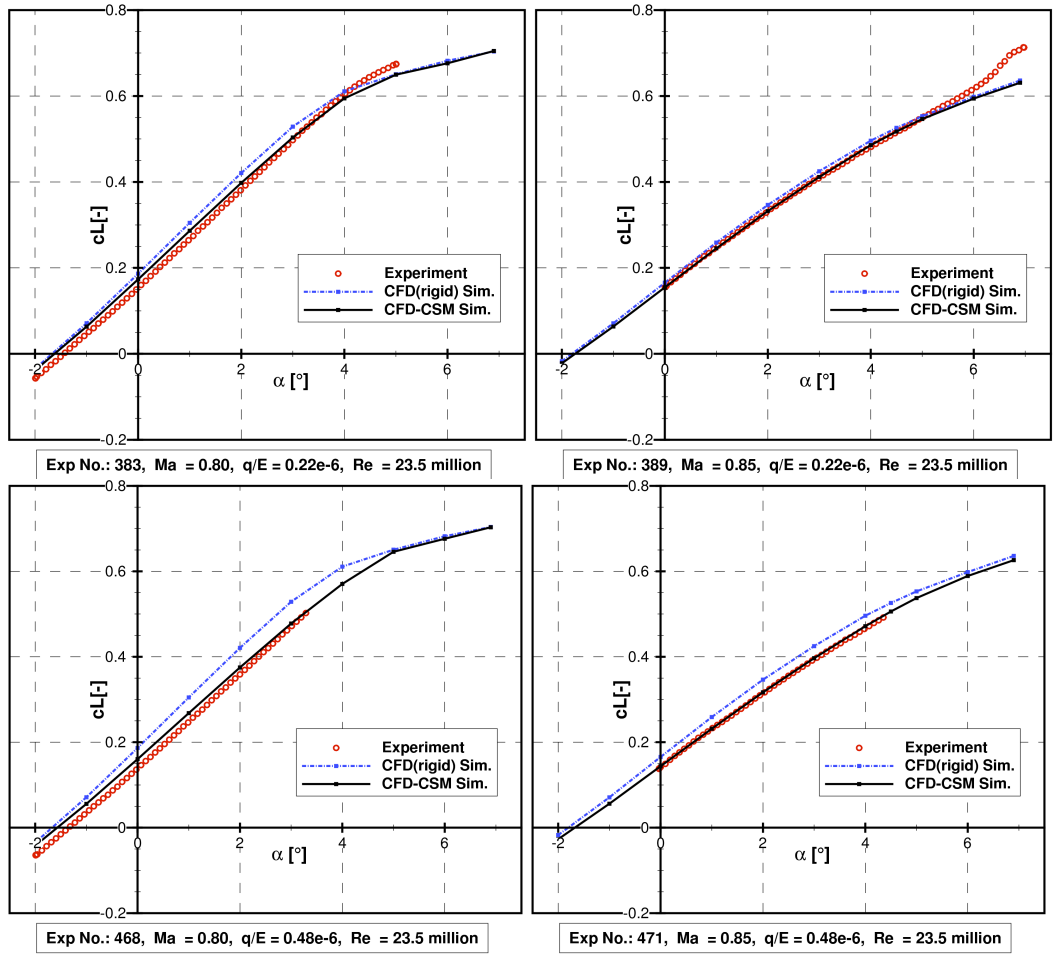

Figure 3, where the cL($\alpha$)-curve for two Mach numbers (0.80, 0.85) and three values of q/E ($0.22 \cdot 10^{-6}$, $0.34 \cdot 10^{-6}$, $0.48 \cdot 10^{-6}$) is presented. The increasing difference between the unchanging data of the rigid assembly computations to the data of the elastic assembly computations is obvious for both Mach numbers with increasing loading factor. Both figures 3 and 9 show a very good agreement of the global lift coefficient between the numerical and experimental data.

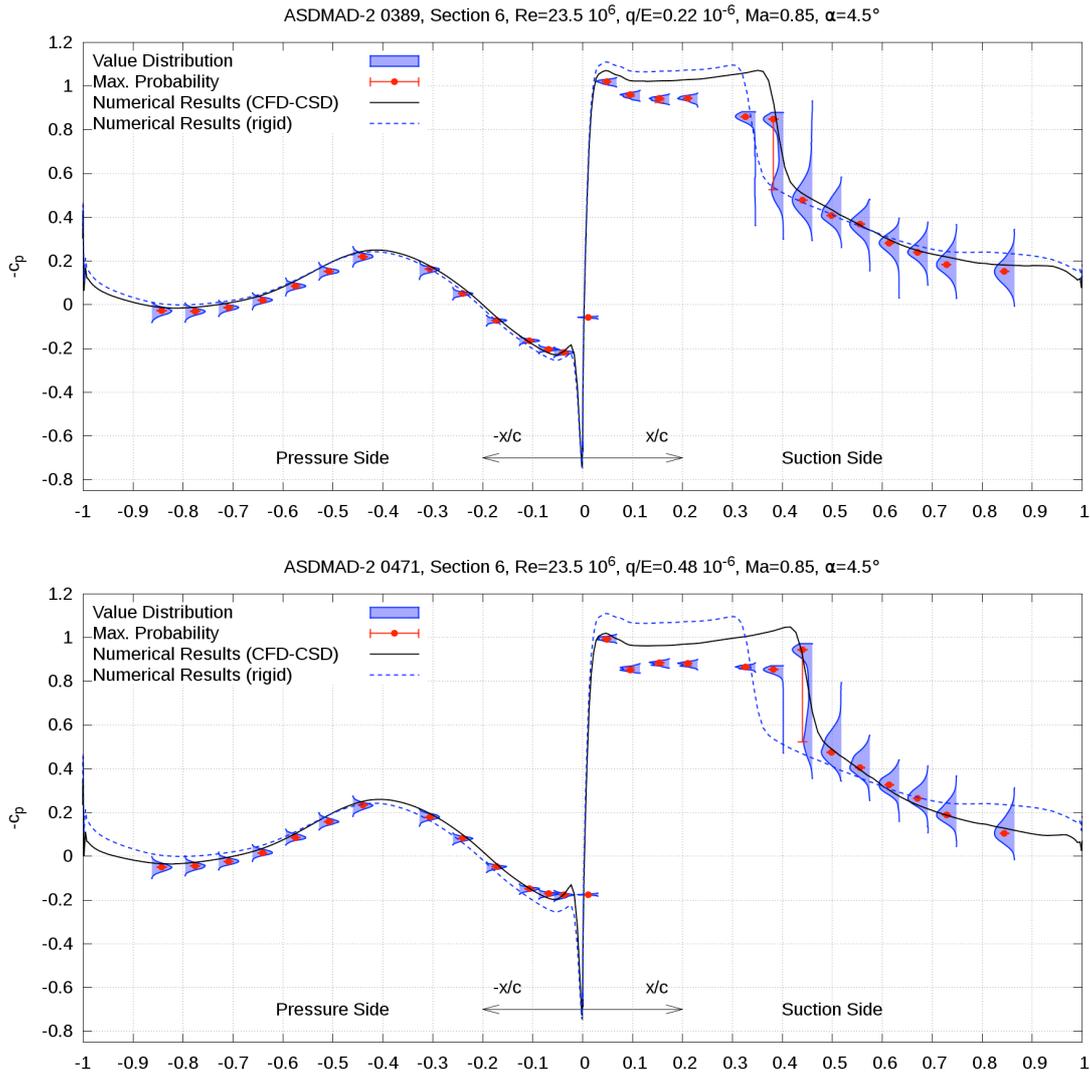

**Figure 9:** Pressure distribution for increasing load factor in section 6, $\alpha=2°$, Ma=0.85 q/E=0.22·$10^{-6}$ (top) and 0.48·$10^{-6}$ (bottom)

## 5 PRELIMIARY UNSTEADY RESULTS

In this section experimental data of the ASDMAD 1 wing (with the one-part winglet) is used as a validation basis. During the unsteady experiments the wing was excited applying inner force couples created by 4 piezo stacks at the wing root. The frequency during this excitation was close to the resonance frequency of the first bending mode to achieve maximum deformation.

Two different simulation approaches are compared in this section:

1. Prescribed Motion Simulation

Because of the excitation close to resonance one may assume that the wing vibrates predominantly in the corresponding mode shape and the simulation could be simplified as a prescribed harmonic motion in that mode shape around the static aeroelastic equilibrium with the amplitude as the single unknown. Since there is no transient response, this approach is computational much less expensive. But it is more or less a pure unsteady aerodynamics case because the mutual reactive effect between solid and fluid are partly disregarded in the computation.

2. Aeroelastic (CFD-CSM) Simulation

Starting from the static aeroelastic equilibrium configuration the wing is excited with sinusoidal oscillating inner force couples, as applied in the experiment, until the oscillation reached a steady state as depicted in Figure 10. As the excitation frequency the simulated free vibration decay frequency (with aerodynamic forces) of corresponding generalized coordinate is used.

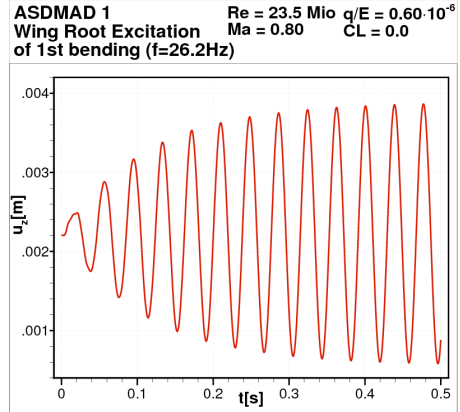

Figure 10: Displacement at the position of the reference acceleration sensor during the transient phase of the wing root excitation

The objective of this section is to investigate to what extent the simulation of the wing root excitation can be simulated in the simplified manner as prescribed harmonic motion.

**5.1 Computational Setup**

For the unsteady results presented here a different coupling chain also based on the FlowSimulator with TAU as flow solver is used. The computation of the structural deformation also uses a modal approach, here with 30 modes. The mode shapes and eigenfrequencies were determined in a preprocessing step with NASTRAN. A loose coupling scheme is used for temporal coupling and a radial basis function based interpolation for the spatial coupling. The CFD mesh was generated with SOLAR and identical boundary conditions, but mostly hexahedrons are used in the vicinity of the no-slip walls. A tetrahedral volume model that also includes wing, clamping, excitation mechanism, balance and the ETW adapter is used as structural model as suggested by Reimer [2].

**5.2 Results**

Simulation and experiment are compared via the Fourier analyzed unsteady pressure data. An acceleration sensor located at the transition between wing and winglet was used as the reference signal.

Figure 11 shows the comparison between the prescribed motion simulation, CFD-CSM simulation and the experimental results for the outermost wing pressure section S6. The 1st bending was excited at two different loading factors $q/E=0.22 \cdot 10^{-6}$ (left) and $q/E=0.60 \cdot 10^{-6}$ (right) for cL = 0.0, Ma = 0.8. These conditions reveal no difference in $cp_{mean}$ and almost no difference in $cp_{amp}$ between the prescribed motion and the CFD-CSM results. On top and bottom an almost constant phase shift of the two different simulation approaches is visible which increases with the load factor. The reason for difference between both simulation approaches is the aerodynamic coupling between the modes, which increases with the load factor. This cannot be considered in the prescribed motion computations. However the difference is small and the prescribed motion computation is capable of capturing the experiment data quite well. One reason for this is that the test assembly and particularly the wing model is very stiff, with the consequence that the natural mode frequencies of the wing in vacuum differ only little form the natural frequencies under wind. Simulation and experiment show a very good agreement in amplitude and phase. The CFD-CSM results agree slightly better than the prescribed motion results with the experimental results.

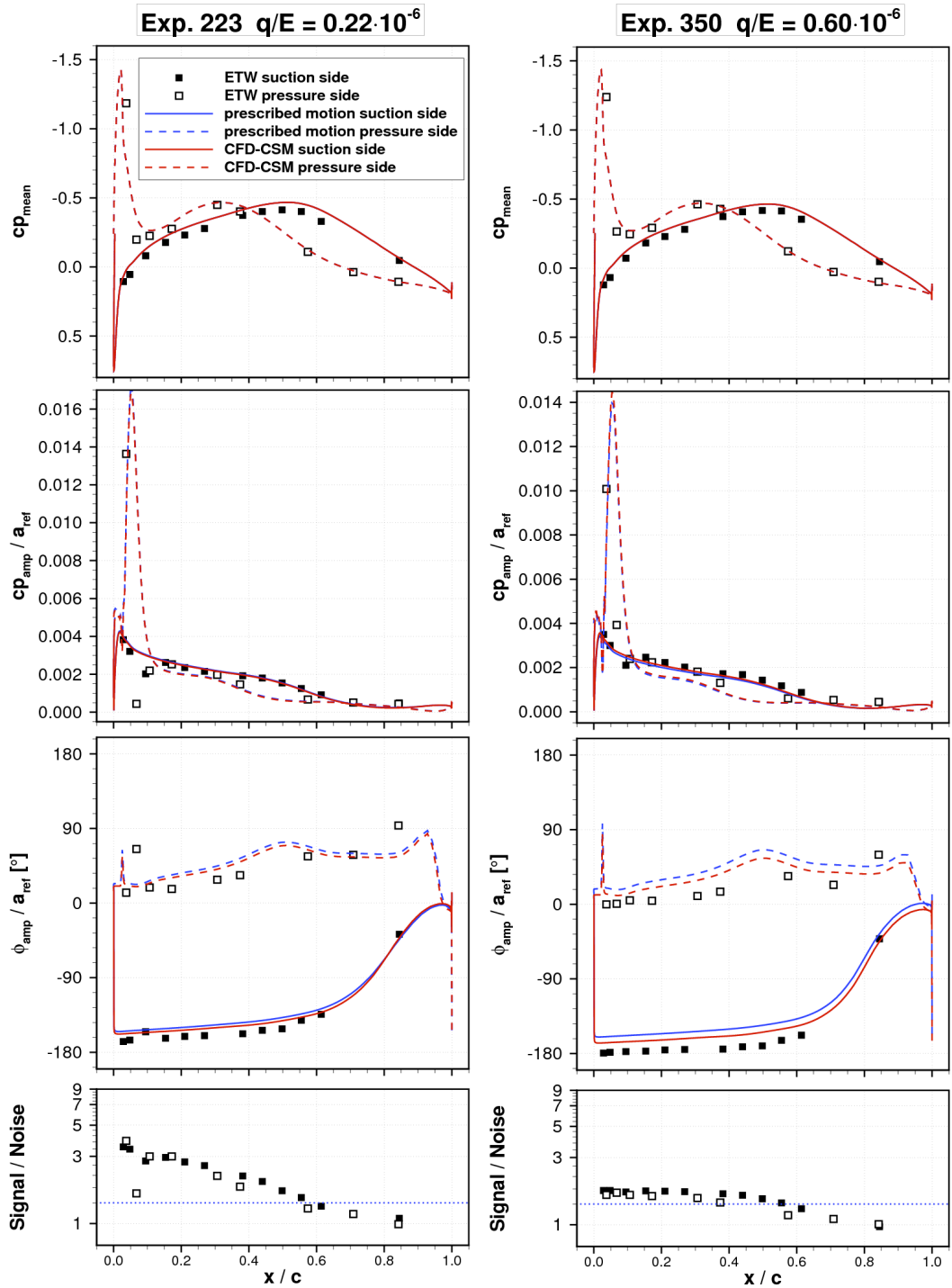

**Figure 11:** Comparison of the Fourier analyzed unsteady cp distribution between CFD-CSM, prescribed motion and experimental results for two different load factors $q/E=0.22 \cdot 10^{-6}$, $0.60 \cdot 10^{-6}$, Ma=0.80, CL=0

## 6 CONCLUSION AND OUTLOOK

In this paper comparisons of results from numerical simulations applying the modular CFD-CSM code SOFIA with TAU as flow solver with selected static wind tunnel experiments of ASDMAD 2 were presented. The dependence of lift on angle of attack, loading factor and Mach number was very accurately captured. With exception of a small offset in front of the shock on the pressure side, the agreement of cp at the pressure sections was very good for all shown variations of angle of attack, loading factor and Mach number. Even the points of maximum probability of cp after strong shock-induced flow separations at high Mach numbers and angle of attacks were reproduced quite well. First numerical and experimental results of unsteady ASDMAD 1 wing root excitation trials were shown. The numerical results were obtained with a slightly different coupling chain that also used TAU as flow solver. To achieve maximum deformations the excitation was chosen close to resonance with natural modes. A comparison of prescribed harmonic motion in the corresponding mode with a true aeroelastic (CFD-CSM) approach where the excitation was realized with force couples at the wing root as in the experiments, revealed no vital difference because the excitation is close the resonance and the aerodynamic coupling between the modes is weak due to the high stiffness of the wing.

Further investigation will focus on the unsteady ASDMAD 2 test, where not only the wing was excited by the wing root excitation mechanism but also by the control surface integrated in the winglet. With additional simulations of the static experiments using DES turbulence modeling the influence of the often-observed shock-induced flow separation to the aeroelastic behavior should be observed. A comparison to the experimental data should show, if these models can confirm the experimental spectrum of the cp value distribution.

## 7 ACKNOWLEGMENT

The funding of this work by the German Research Foundation and Airbus Operations GmbH within the framework of the transfer project ASDMAD is gratefully acknowledged. Computing resources were provided by the RWTH Aachen University Center for Computing and Communication and supported by the German Research Foundation under GSC 111 (AICES). The unsteady results were obtained in cooperation with Airbus within the ASDMAD project.